\documentclass[12pt]{article}

\usepackage{amsmath,amssymb, amsthm}

\title{V\"axj\"o Interpretation of Wave Function: 2012}

\author{Andrei Khrennikov\\International Center for Mathematical Modelling\\
in Physics and Cognitive Sciences\\ Linnaeus University, V\"axj\"o,
S-35195, Sweden}

\begin{document}

\maketitle

\begin{abstract}
We discuss the problem of interpretation of the wave function. The latest version of the V\"axj\"o 
interpretation is presented. It differs essentially from the original V\"axj\"o 
interpretation (2001).  The main distinguishing feature of the present V\"axj\"o 
interpretation is the combination of realism on the subquantum level with nonobjectivity
of quantum observables (i.e., impossibility to assign their values before measurements). 
Hence, realism is destroyed by detectors transforming continuous subquantum reality into 
discrete events, clicks of detectors. The V\"axj\"o 
interpretation-2012 is fundamentally contextual in the sense that the value of an observable 
depends on measurement context. This is contextuality in Bohr's sense. It is more general 
than Bell's contextuality based on joint measurements of compatible observables.   
\end{abstract}

\maketitle

\section{Introduction}

During the V\"axj\"o series of conferences on quantum foundations the most exciting spectacle started each time when the
question of {\it interpretations of the wave function}  \index{interpretation of wave function} attracted
the attention. Finally, I understood that the number of different
interpretations is in the best case equal to the number of
participants. If you meet two people who say that they are
advocates of, e.g., the Copenhagen interpretation of quantum mechanics (QM), ask them
about the details. You will see immediately that their views on what is
the Copenhagen interpretation can differ very much. The same is
true for other interpretations. If two scientists tell that they
are followers of Albert Einstein's {\it ensemble interpretation,} \index{ensemble interpretation}
ask them about the details... At one of the round tables (after two hours of
debates with opinions for and against completeness of QM) we had
decided to vote on this problem. Incompleteness advocates have won,
but only because a few advocates of completeness voted for
incompleteness. The situation is really disappointing: the basic
notion of QM has not yet been properly interpreted (after 100
years of exciting, but not very productive debates).\footnote{``When
I speak with somebody and get to know their interpretation, I
understand immediately it is wrong. The main problem is that I do
not know whether my own interpretation is right.'' (Theo Nieuwenhuizen) \index{Nieuwenhuizen}This is the
standard problem of participants of V\"axj\"o conferences.} I specifically
appreciate the activity of Arkady Plotnitsky, \index{Plotnitsky}
 philosopher studying Bohr's views, see, e.g., \cite{P4}. He
 teaches us (participants of V\"axj\"o conferences) about interpretations of the wave functions a lot. First of
 all we got to know that the Copenhagen interpretation \index{Copenhagen interpretation} cannot be
 rigidly coupled with Bohr's views. On many
 occasions  Niels Bohr emphasized that QM is not about physical processes in microworld,
 but about our measurements \cite{BR}:

 ``Strictly speaking, the mathematical formalism of quantum mechanics and electrodynamics
merely offers rules of calculation for the deduction of expectations pertaining
to observations obtained under well-defined experimental conditions specified
by classical physical concepts''.

The basic postulate of the Copenhagen
 interpretation of QM  -- ``the wave function describes completely the state
 of a quantum system'' (i.e., a concrete system, not an ensemble)
 -- cannot be assigned to Bohr. Then we learned (again from
 Plotnitsky)  that Bohr's views have been crucially changed a few times
 during his life. Thus, there can be found many different Bohr's
 interpretations of QM. Bohr was definitely the  father of the {\it operational
 interpretation} of QM.  As was already pointed out, Bohr emphasized that the formalism
of QM does not provide the intrinsic description of processes in microworld, it describes only results
of measurements.   Bohr also can be considered as
 one of fathers of the so-called {\it information interpretation} \index{information interpretation} of
 QM: the QM-formalism describes information about micro systems
 extracted by means of macroscopic measurement devices. Heisenberg
(and to some extent Schr\"odinger) shared this viewpoint. Nowadays
the information interpretation of QM became very popular, see,
e.g., \cite{KV02}, \cite{Folse}, \cite{P4}. I can mention Anton Zeilinger \index{Zeilinger} \cite{Zeilinger} and Christopher
Fuchs  \index{Fuchs } \cite{Fuchs}--\cite{Fuchs1} among the active promoters of this interpretation; we can also mention Mermin's paper \index{Mermin} \cite{ME1}.

In this paper we discuss three main interpretations of QM: {\it 
the Copenhagen, ensemble and V\"axj\"o interpretations.}

\section{Interpretations of Wave Function}

Everywhere below $H$ denotes complex Hilbert space with the
scalar product $\langle \cdot, \cdot \rangle$ and the norm $\Vert
\cdot \Vert$ corresponding to the scalar product.

\medskip

{\bf  Postulate IM.} (The mathematical description of quantum
states). {\it  Quantum (pure) states (wave functions) are
represented by normalized vectors $\psi$ (i.e.,
$\Vert\psi\Vert^2=\langle \psi,\psi \rangle =1$) of a complex
Hilbert space $H.$ Every normalized vector $\psi \in H$ may
represent a quantum  state. If a vector $\psi$ corresponding to a
state is multiplied by any complex number $c, \vert c \vert =1,$
the resulting vector will correspond to the same
state.}\footnote{Thus states are given by elements of the unit
sphere of the Hilbert space $H.$}

\medskip

The physical meaning of ``a quantum state'' is not defined by this
postulate. It must be provided by a separate postulate; see
Postulates IE, IC, IV (respectively, the ensemble, Copenhagen and V\"axj\"o interpretations). 

\medskip

\medskip

{\bf Postulate IE.} (The ensemble  interpretation). {\it A wave
function provides a description of certain statistical properties of
an ensemble of similarly prepared quantum  systems.}

This interpretation is upheld, for example by Einstein, Popper,
Blokhintsev, Margenau, Ballentine, Klyshko, and recent years by,
e.g.,  De Muynck, De Baere, Holevo, Santos, Khrennikov,
Nieuwenhuizen, Adenier, Groessing, Hofer and many others 

\medskip

{\bf Postulate IC.} (The Copenhagen interpretation).
{\it A wave function provides a complete description of an
individual quantum system.}

\medskip

This interpretation was supported by a great variety of members,
from Schr\"odinger's original attempt to identify the electron
with a wave  function solution of his equation to the several
versions of the Copenhagen interpretation; for example,
Heisenberg, Bohr, Pauli, Dirac, von Neumann, Landau, Fock, ..., 
Greenberger, Mermin, Lahti, Peres.

There is an interesting story about the
correspondence between Bohr and Fock on the individual
interpretation. This story was told me by one of former students
of Fock who pointed out that one of the strongest supporters of
this interpretation was Vladimir A. Fock, and that even though
Bohr himself had doubts about its consistency, he, Fock,
demonstrated to Bohr inconsistency in the Einsteinian ensemble
interpretation. Thus interpretation which is commonly known as the
{\it Copenhagen interpretation} might be as well called the ``{\it
Leningrad interpretation}.'' 

\medskip

Instead of Einstein's terminology "{\it ensemble interpretation}",\index{ensemble interpretation}
L. Ballentine \cite{BLA}, \cite{BL} used the terminology {\it
``{\it statistical interpretation}.''}\index{statistical interpretation} However, Ballentine's
terminology is rather misleading, because the term ``statistical
interpretation'' was also used by J. von Neumann \cite{VN} for individual
randomness! For him ``statistical interpretation'' had the meaning
which is totally different from the Ballentine's
``ensemble-statistical interpretation.'' J. von Neumann wanted to
emphasize the difference between deterministic (Newtonian)
classical mechanics in that the state of a system is determined by
values of two observables (position and momentum) and quantum
mechanics in that the state is determined not by values of
observables, but by probabilities.  We shall follow Albert
Einstein and use the terminology {\it ``ensemble
interpretation''.}

\section{V\"axj\"o interpretation of quantum mechanics}

The V\"axj\"o interpretation\index{V\"axj\"o interpretation} \cite{Vaxjo1} was born (in 2001) as an alternative to the Copenhagen interpretation.
The basic assumption of the latter that the wave function describes completely
the state of a quantum system is the main source of all quantum mysteries. (We remind that Schr\"odinger elaborated  the``Schr\"odinger cat''\index{Schr\"odinger cat}
experiment to show the absurdness of this assumption. Nowadays the origin of the 
Schr\"odinger cat illustration of the absurdness of the Copenhagen interpretation is practically forgotten. The presence of such cat-states is often considered as one of natural features of QM.) Therefore the easiest way to resolve the majority of interpretation problems of QM is to assume that the wave function description is not the final description of micro phenomena. By proceeding in this way one has to develop prequantum models inducing the QM formalism as an operational formalism
ignoring details of processes in the microworld and describing  only results of measurements. Measurements are operationally described by Hermitian operators or more generally by POVM. However, there are known many no-go statements which prohibit any motion beyond QM (under the natural assumption of locality\footnote{We remind that Einstein, Podolsky and Rosen considered nonlocality as the {\it absurd alternative} to incompleteness of QM \cite{EPR}. This is also practically forgotten (ignored?). One can often find just the statement that quantum nonlocality was originally discussed by Einstein, Podolsky and Rosen \cite{EPR}. In his reply to the EPR paper \cite{BR0} Bohr did not mention nonlocality at all. He also was sure that QM is local.}).  The first version \cite{Vaxjo1} of the V\"axj\"o interpretation was created as the result of analysis of the ``impossibility statements'', e.g.,  \cite{VN}, \cite{KH}, \cite{B}, playing the crucial role in the contemporary debates on quantum foundations. Surprisingly for myself, I found that {\it all no-go statements} contain some unphysical assumptions which are not valid for real experimental situations, see \cite{KHR-context} for details; cf. with Bell's and Ballentine's critical analysis of assumptions of von Neumann no-go theorem \cite{B}, \cite{BL}, \cite{BLA}. Suddenly I understood that usage of the operational quantum formalism for results of measurements does not contradict with the possibility of creating prequantum classical models. These models can be both realistic and local. Here {\it realism (objectivity) is defined as a possibility to assign the values of quantum observables to quantum systems before measurement.}\index{realism}\index{objectivity}
This viewpoint to realism is common in discussions related to Bell's inequality \cite{B} and in general to inter-relation of classical and quantum physics. 

\medskip

{\it The V\"axj\"o interpretation-2001 was the (local) realistic interpretation \cite{Vaxjo1}.}

\medskip

In 2004 I visited Beckman Institute for Advanced Science and Technology (University of Illinois at Urbana-Champaign). After my talk (May 3, 2004), I 
discussed with A. Leggett  a role of no-go theorems in QM. In particular, I wondered why Einstein had never mentioned the von Neumann's no-to theorem. (After Einstein's death, the book of von Neumann \cite{VN} was found in Einstein's office.) A. Leggett remarked that Einstein was mainly interested
in the real physical situation rather than in formal mathematical statements. I started to think about the real physical situation described by QM.  I understood that {\it even if some classical prequantum model is formally possible, in spite of all no-go ``theorems'', this does not imply that this model matches with the real physical situation.} Hence, although local realism cannot be rejected as a consequence of, e.g., violation of Bell's inequality, one has carefully to analyze matching of local realism with the real physical situation. Since, as for Einstein and Bohr, I did not take seriously quantum nonlocality, for me  the questionable point was the possibility of realistic representation of quantum observables.

 By reading Bohr's works \cite{BR0}, \cite{BR} I understood the fundamental role of {\it measurement context}\index{context} in quantum measurements. Bohr stressed that the result of measurement is the sum of impacts of the quantum system and the measurement device. And it is impossible to distill the contribution 
of the system from the integral measurement result. Thus quantum observables are not objective. We could not assign a value of, e.g., position to a quantum system before measurement. We remark that  we discuss measurement of a single quantum observable and not of a pair of observables. Hence, the essence of this discussion is not in the impossibility of the joint measurement of some quantum observables, so called incompatible observables, e.g., position and momentum (the Bohr's principle of complementarity \cite{BR}). We discuss {\it contextuality}\index{contextuality} of even a single quantum observable: the properties of measurement context cannot be separated from intrinsic properties of a quantum system.\footnote{  May be Bohr would not
accept that a quantum system has intrinsic properties at all. (Bohr's works are very difficult for understanding; try to read , e.g., \cite{BR0}.) At least it is clear that he was not an idealist. So,   he might agree that, e.g., an electron has some intrinsic properties.
However, for Bohr, these properties were fundamentally inapproachable in measurements.} 

Unfortunately, in recent discussions related  to Bell's theorem \cite{B} the notion of contextuality is coupled to the joint measurement of compatible observables. In this book (following  N. Bohr), we always use  the notion of contextuality as  the irreducible dependence of the result of measurement (of even a single observable) on the whole experimental arrangement. 

Under influence of Bohr I rejected {\it Einstein's realism.}\footnote{I also mention an intermediate version of the V\"axj\"o interpretation \cite{KLA} in which I tried to combine realism with contextuality, see also \cite{KHR-context}. This interpretation might be useful in some applications of quantum information outside of physics.} Here two circumstances of totally different nature were very important. 

One of my main scientific interests 
is the application of the mathematical formalism of QM outside of physics: to study statistical data from cognitive science, psychology, finances \cite{KHR-finance}. In these applications observables, ``mental observables'', are nonobjective by their nature; contexts play a crucial role. And this is well known for psychologists and researchers working in cognitive science, sociology, economics. This out-physics activity improved my understanding of the role of contextuality. 

Another strong motivation to reject realism of quantum observables\footnote{We state again  that in this discussion realism is regarded as objectivity of quantum observables, i.e., the possibility to assign the value of a quantum observable to the quantum   
 system {\it before measurement.} We do not reject realism in its general philosophical meaning. In particular, quantum systems have their own properties, properties of objects. However, these objective properties could not be approached by the measurement devices in use. The problem is that the class of measurement devices is not large enough to approach the subquantum level. Of course, this theoretical discussion does not imply that it would be possible (at least for our civilization) to elaborate novel measurement devices to approach subquantum spatial and temporal scales.} was related to the development of a measurement theory for 
{\it prequantum classical statistical field theory}, PCSFT \cite{KH1}--\cite{KHAIP10}. This is the classical field theory which reproduces quantum averages,
including correlations for quantum observables (e.g., spin or polarization) for entangled systems. Thus, opposite to claims of adherents
of the Copenhagen interpretation (starting with von Neumann \cite{VN}), quantum randomness can be reduced to randomness of 
classical systems. 
There is, however, a crucial proviso. While creation of PCSFT implies QM can be interpreted as a form of classical statistical mechanics,
this classical statistical theory is not that of particles, but of fields. This means that the mathematical formalism of QM must be translated into the mathematical formalism of classical statistical mechanics on the infinite-dimensional phase space. The infinite dimension of the phase space of this translation is a price of classicality. From the mathematical viewpoint this price is very high, because in this case the theories of measure, dynamical systems, and distributions are essentially more complicated than in the case of the finite-dimensional phase space found in classical statistical mechanics of particles.\footnote{However, at the model level (similar to quantum information theory) one can proceed with the finite-dimensional phase space
by approximating physical prequantum fields by vectors with finite number of coordinates.} 
On the other hand, from the physical and philosophical viewpoints, considering QM as classical statistical mechanics of fields can resolve the basic interpretational problems of QM. As was already pointed out, for example and in particular, quantum correlations of entangled systems can be reduced to correlations of classical random fields. From this perspective, quantum entanglement is not mysterious at all, since quantum correlations are no longer different from the classical ones.

However, usage of PCSFT reduces quantum randomness to classical only on the level of averages and not  individual events,
clicks of detectors. PCSFT is the theory of classical  random fields reproducing quantum averages (including correlations for entangled
systems) as averages of quadratic forms of fields. Since the basic model of  prequantum fields is the Gaussian one, these fields are continuous. Hence, in the same way
as in the classical signal theory, averages are calculated for variables with a {\it continuous} range of values. This model, PCSFT, is really a {\it prequantum} model. In accordance
with the Bohr's viewpoint, we consider QM as an operational formalism describing (predicting) results of measurements on micro systems. QM cannot describe intrinsic physical processes
in the microworld, but only measurements performed by macroscopic classical devices. In contrast, PCSFT describes intrinsic processes in the microworld. However, it does not describe
results of measurements. In principle one might be satisfied with creation
of a prequantum model which reproduces only quantum averages without establishing a direct connection with  theory of measurement on the level of individual events. This approach would match with views of Schr\"odinger and the {\it Bild concept} in general.
(The Bild concent was formulated by Hertz
in his 1894 {\it Prinzipien der Mechanik.}  Schr\"odinger's views to QM were based on this concept.) 
However, historical development of QM demonstrated that the Bild concept was not attractive for
the majority of physicists. Since the experimental verification is considered as the basic counterpart of any physical theory, Bild-like prequantum theories are considered as metaphysical.
Measurement theory connecting PCSFT with experiment on the level of individual events was developed in  my paper \cite{Tr1}.
In this paper I elaborated a scheme of discrete measurements of classical random signals which reproduces the basic rule of QM, {\it the Born's rule.} This is the scheme of  {\it threshold type detection}\index{threshold detection}: such a detector clicks after it has ``eaten''
a special portion of energy $\epsilon$  of a prequantum random signal.  We call such measurement theory threshold detection (TSD) model.
TSD is a classical measurement model which delivers the same predictions as QM.  In contrast to QM, which is an operational formalism\footnote{In QM measurement devices are simply black boxes which are symbolically represented by self-adjoint operators or more generally (and even more formally) by positive operator  valued measures (POVM). }, TSD provides a detailed description of the energy balance in the process of interaction of a classical random field (fluctuating at a fine time scale) with a detector, as well as the conditions inducing an individual quantum event, a click. 
Thus QM can be interpreted as theory of measurements of classical
random signals with the aid of threshold type detectors. They produce discretization of continuous classical fields. This discretization is the essence of quantum phenomena. Hence,
``quantumness is created in detectors", there is no quantumness without measurement. In particular, this viewpoint contradicts the views of
early Einstein who claimed that the electromagnetic field is quantized not only in the process of the energy exchange with material bodies, but even in vacuum. In the  PCSFT/TSD model photons appear only on the level of detection, so they are nothing else than the clicks of detectors. Surprisingly this viewpoint on the notion of photon
coincides with the views of some top level experimenters working in quantum optics, e.g., A. Zeilinger\index{Zeilinger}, A. Migdal\index{Migdal}, and S. Polyakov\index{Polyakov} also identify photon with a click of a detector (their talks at V\"axj\"o conferences). Of course, the majority
of experts in quantum optics (including aforementioned) proceed under the Bohr's assumption on completeness of QM.

Our threshold detection model is fundamentally based on the {\it assumption of ergodicity}\index{ergodicity} of prequantum random signals, i.e., the time and ensemble averages coincide.
We stress that this ergodicity\index{ergodicity} is related to the subquantum processes, i.e., to the intrinsic physical processes in the microworld. Another important assumption
is that the prequantum random signals fluctuate on a  very fine time scale (comparing with the time scale of lab-measurements). The duration of measurement, i.e., the interaction
of a random field with a detector, is huge on the prequantum time scale. However, it is sufficiently small with respect to the time scale of lab-measurements.
(The rigorous mathematical justification of the derivation of the Born's rule in the TSD-framework is rather complicated mathematically and it is based on the theory of {\it ergodic stochastic processes.})

We remark that ergodic processes are {\it  stationary.} In the process of interaction with a detector, a prequantum field (emitted by a source and evolving in the free space in accordance with the Schr\"odinger equation with a random initial condition approaches a steady state, the field becomes stationary. This first phase is  very short. Then the
stationary stochastic process transmits its energy to the detector and this stage is very long (infinitely long in the pure mathematical framework). We remark that this discussion is about the prequantum time scale.

Max Born\index{Born} postulated the probabilistic rule coupling the wave function
 with quantum detection probabilities:
the probability equals to the squared wave function.
Whether it is possible to derive this rule from some natural physical principles is still the subject of intensive
debates. In \cite{Tr1} we  derived the Born's rule in the PCSFT/TSD framework: threshold type detectors interacting with random signals of a special class produce a statistics of clicks which is (approximately)  described by the
Born's rule. Of course, this framework is based on PCSFT-coupling between the covariance operator of a random field and the density operator of the quantum system corresponding to this random field, see  \cite{KH1}--\cite{KHAIP10} for details.\footnote{ This derivation of the fundamental rule of QM which establishes the basic relation of the theoretical predictions of QM with experiment definitely would not match with the expectation
of those who consider QM as theory which is full of mysteries.}

PCSFT/TSD also proposed very natural solution of quantum {\it measurement problem} \cite{VN}. Continuous evolution of the prequantum random field induces discrete jumps corresponding
to approaching the detection threshold. Such jumps are formally described by the von Neumann projection postulate.  However, this solution of the measurement problem is also
far from the expectations of people excited by quantum exotics. Moreover, since, to solve the measurement problem, we go beyond QM, our solution would not be considered as the
solution of the problem formulated by von Neumann \cite{VN}. The latter is expected to be solved in the conventional quantum framework.

However, the TSD-transition from (continuous) prequantum random fields to clicks of detectors (discrete events) destroys the 
PCSFT-objectivity. A prequantum field  emitted by a source does not predetermine the result of measurement.
There are a few different sources of non-objectivity of quantum observables:

{\bf BF}. The presence of the irreducibly random background field. 
PCSFT can reproduce quantum correlations for entangled systems only by taking into account the random background.
Its presence (everywhere in space) makes correlations stronger than they could be expected. In particular, by PCSFT 
Bell's inequality is violated only due to the contribution of the random background. Hence, by emitting the concrete pulse from 
a source we are not able to predict its structure at the moment of arrival to a detector; moreover, the random background is also 
present in detectors. Therefore we are not able to predict the output of the detector, even if we were able to eliminate the 
random background in space between the source and the detector.     However, this is simply the well known problem of classical stochasticity. As in classical physics, we may idealize the situation and to assume that the random background can be eliminated
(although in reality this is definitely impossible).  Nevertheless, objectivity cannot be recovered even under
this assumption, see {\bf CONT} and {\bf DC}.

 {\bf CONT}. In TSD (in the total accordance with Bohr's views) the result of a measurement depends fundamentally 
on experimental context, on all parameters of the detector in use (and, in particular, on the threshold in  the detector).
For the same classical field, detector can click or not depending on its settings. On the level of individual events such ``properties'' as spin and polarization cannot be assigned to the prequantum wave.   They are determined by detection context.
(We remark that a prequantum wave definitely has such properties, but in continuous and not discrete representation).
Thus PCSFT/TSD is contextual, in general Bohr's and not restricted Bell's sense. Suppose now that we fixed all 
parameters of a detector. Can we speak about objectivity of the corresponding quantum observable?  (Of course, under the assumption
that the random background was totally eliminated.) Still  not, see {\bf DC}.

{\bf DC}.  In QM coincidence counts of two detectors reacting to a single system emitted by a source are interpreted as 
the artifacts induced by noise. In TSD such coincidence counts are fundamentally irreducible. This is a consequence of the 
threshold detection scheme. Hence, real observables are multi-valued. Even the concrete prequantum wave interacting with 
the concrete detector and in the total absence of the random background can produce clicks in two detectors for, e.g., 
 measurement of spin projection to two axes. It is impossible to assign to this wave neither the value spin up nor spin down.

The new V\"axj\"o interpretation combines Einstein's and Bohr positions. As well as Einstein, we do not believe that the quantum state describes completely the physical state of a quantum system.  The quantum state describes only statistical features. As well as Bohr, we do not consider quantum observables as objective quantities (intrinsic properties of objects). Quantum observables are fundamentally contextual.

\medskip

{\bf Postulate IV.} (The V\"axj\"o interpretation).
{\it A wave function describes correlations in prequantum random 
fields (which are symbolically represented as quantum systems).}

\medskip

This paper was written under support of the grant "Mathematical Modeling of Complex Hierarchic Systems" of 
the Faculty of Natural Science and Engineering of Linnaeus University.

\end{document}